\documentclass[twocolumn,english,aps,superscriptaddress,prb]{revtex4-2}
\usepackage{graphicx}
\usepackage{amsmath,xcolor}
\usepackage{hyperref}
\usepackage{soul}

\newcommand{\ie}{\textit{i.e., }}
\newcommand{\pd}{{\phantom{\dagger}}}

\newcommand{\bq}{\boldsymbol{q}}
\newcommand{\bK}{\boldsymbol{K}}
\newcommand{\br}{\boldsymbol{r}}

\DeclareUnicodeCharacter{2212}{\textendash}

\begin{document}

\title{Chiral topological superconductivity in hole-doped Sn/Si(111)}

\author{Matthew Bunney}
\affiliation{School of Physics, University of Melbourne, Parkville, VIC 3010, Australia}
\affiliation{Institute for Theoretical Solid State Physics, RWTH Aachen University, 52062 Aachen, Germany}
\author{Lucca Marchetti}
\affiliation{School of Physics, University of Melbourne, Parkville, VIC 3010, Australia}
\author{Domenico Di Sante}
\affiliation{Department of Physics and Astronomy, University of Bologna, 40127 Bologna, Italy}
\author{Carsten Honerkamp}
\affiliation{Institute for Theoretical Solid State Physics, RWTH Aachen University, 52062 Aachen, Germany}
\author{Stephan Rachel}
\affiliation{School of Physics, University of Melbourne, Parkville, VIC 3010, Australia}

\begin{abstract}
A third monolayer of tin atoms on the semiconductor substrate Si(111) has been shown to become superconducting upon six to ten percent hole doping. Experiments have reported promising results hinting at a superconducting chiral $d$-wave order parameter. Here we examine Sn/Si(111) by combining most recent ab initio results, quasi-particle interference calculations, state-of-the-art truncated-unity functional renormalization group simulations and Bogoliubov-de Gennes analysis. We show remarkable agreement between experimental and theoretical quasi-particle interference data both in the metallic and superconducting regimes. The interacting phase diagram reveals that the superconductivity is indeed chiral $d$-wave with Chern number $C=4$. Surprisingly, magnetically ordered phases are absent, instead we find charge density wave order, as observed in related compounds, as a competing phase. Our results demonstrate that Sn/Si(111) is an outstanding candidate material for chiral topological superconductivity.
\end{abstract}

\date{\today}

\maketitle

Topological superconductors are one of the most fascinating states of matter -- even though they are superconductors, they possess protected edge states analogous to (topological) Chern insulators. Indeed, a chiral topological superconductor can be considered as a Chern insulator of superconducting quasiparticles. At point-like defects in the material, these quasiparticles are localized Majorana bound states, highly speculated to be components of next-generation quantum technology\,\cite{nayak-08rmp1083,elliott-15rmp137,fu-10prl056402}.

There have been many candidate materials for topological superconductors\,\cite{kallin2016,sato2017,mandal2023}, but few have stood the test of time\,\cite{maeno1994, maeno2024}. Current candidates include  UTe$_2$\,\cite{ran2019, jiao2020, hayes2021}, Fe(Se,\,Te)\,\cite{machida2019}, heavy fermion systems such as UPt$_3$\,\cite{stewart1984, schemm2014}, rhombohedral graphene\,\cite{han2025}, transition metal dichalcogenides such as WS$_2$ and MoTe$_2$\,\cite{hsu2017}, kagome metals AV$_3$Sb$_5$ (A = K, Rb, Cs)\,\cite{deng2024, le2024}, and engineered platforms such as non-centrosymmetric superlattices\,\cite{wan2024}, and most recently also Sn/Si(111)\,\cite{ming2023}.

Despite only recent claims of topological superconductivity\,\cite{ming2023}, Sn/Si(111) is a rather old material, first proposed and synthesized in 1964\,\cite{estrup1964,kinoshita1986}. The Sn adatoms form a 1/3 depleted surface layer relative to the bulk, which garnered significant interest as the half-filled surface band forms a quasi-two dimensional system from the `dangling bonds' of the Sn atoms, which is energetically isolated from the bulk bands\,\cite{profeta2007,li-11prb041104,li2013,hansmann2013,marchetti2025}.
The 6\%--10\% hole-doped system is metallic, before entering the superconducting regime at low-temperatures\,\cite{ming2023}.
In contrast, the half-filled system is known to feature a small gap down to 5K in the differential tunneling conductance of scanning-tunneling spectroscopy  as well as in angle-resolved photoemission experiments\,\cite{modesti2007,li2013}; measurements below 5K are absent.
The gap at half filling led early on to speculations about a Mott-insulating ground state\,\cite{profeta2007,modesti2007}. Further angle-resolved photo emission spectroscopy measurements were interpreted as indirect evidence of row-wise antiferromagnetic order\,\cite{li2013}. Motivated by the analogy to the high-temperature cuprate superconductors, it was proposed that doped Sn/Si(111) should lead to an unconventional superconducting state with $d$-wave symmetry\,\cite{profeta2007,ming2017,cao2018}.

The experimental discovery of superconductivity in hole-doped Sn/Si(111), where the modulation doping of the Si substrate leaves the perfect triangular surface intact\,\cite{wu2020b,ming2017,ming2023}, inspired further theoretical investigation: it was shown that in the infinitesimal-coupling limit the leading instabilities would be spin triplet-superconductivity if non-local Hubbard interactions are taken into account\,\cite{wolf2022c,biderang2022}. Stronger electron-electron interactions were shown to advantage unconventional singlet-superconductivity\,\cite{wolf2022c}, in line with the earlier predictions about $d$-wave singlet superconductivity\,\cite{profeta2007,cao2018}. It was also argued that charge fluctuations are the driving mechanism for superconductivity in Sn/Si(111)\,\cite{hansmann2016,cao2018}.

Current experimental evidence of the chiral superconductivity is based on scanning tunneling microscopy (STM) methods, but while experimental activity has been growing to change this\,\cite{wu2025, nishimichi2025, iannetti2025,iannetti-23arXiv}, the extremely clean samples invite one to directly compare theory and experiment. Since the details of the superconducting pairing affect how it scatters off impurities, quasi-particle interference (QPI) has been put forward as a diagnostic tool for unconventional superconducting pairings\,\cite{capriotti2003, hirschfeld2015, altenfeld2018, boker2020}.

In this Letter, we will take advantage of the sample quality of boron-doped Sn/Si(111)\,\cite{ming2023} and compare the QPI spectra both in the metallic and superconducting regimes with our theory. We use the most recent \textit{ab initio} bandstructure data\,\cite{marchetti2025} as input for many-body truncated-unity functional renormalization group (TUFRG) simulations\,\cite{husemann2009, wang2012, lichtenstein2017} and evaluate the interacting phase diagram. We then analyze these results further within the mean-field framework, allowing us to narrow down the possible ground states of the many-body phases, which turn out to be in excellent agreement with the experimental findings.

\begin{figure}[t!]
    \centering
    \includegraphics[width=\linewidth]{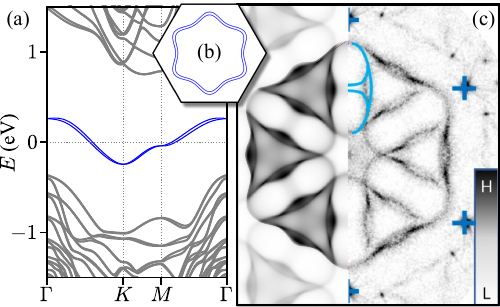}
    \caption{Metallic regime of Sn/Si(111). (a) Band structure of Sn/Si(111), the surface band of the Sn atoms highlighted in blue\,\cite{marchetti2025}. (b) Corresponding Fermi surface of the relevant doping regimes. c) QPI images at 10\% hole-doping in momentum space: (Right) Experimental data\,\cite{ming2023} at $\omega=0$mV, $T=9$K; figure reused with permission of the authors\,\cite{reuse}.
    }
    \label{fig:qpi}
\end{figure}

{\it QPI in the metallic regime}.---We start by using the Wannier model of the metallic surface band of Sn/Si(111)\,\cite{marchetti2025} (see Fig.\,\ref{fig:qpi}\,a and b) to calculate the QPI image in momentum space. Experimentally, such images are obtained by Fourier transforming STM scans over a region containing surface impurities, which contains the distinct scattering patterns of the electronic surface state scattering off the impurity. In Ref.\,\onlinecite{ming2023}, such scans were performed for 6, 8 and 10\% boron-doping below and above $T_c$; the latter corresponds to the metallic regime.

The bandstructure\,\cite{marchetti2025} can be compactly written as
\begin{equation}\label{ham}
    H_0 = \sum_{ij \, \sigma \sigma'} t^{\phantom{\dagger}}_{ij \, \sigma \sigma'} c^\dagger_{i \sigma} c^\pd_{j \sigma'} - \mu \sum_{i \, \sigma} c^\dagger_{i \sigma} c^{\phantom{\dagger}}_{i \sigma} \, ,
\end{equation}
where $i,j$ index the Sn surface adatoms, $\sigma \sigma'$ their spins and $t_{ij \, \sigma \sigma'}$ the tight binding hopping parameters extracted from the Wannier model. They contain real-valued hoppings, spin-dependent hopping terms corresponding to Rashba spin-orbit coupling, radial Rashba coupling and a valley-Zeeman term; the latter two terms are very weak. The second term in \eqref{ham} is the chemical potential which allows an adjustment of the doping $\delta$ with $\delta = 0$ corresponding to half-filling $n=1$. The two split van Hove peaks are at (hole) dopings $\delta = -0.246, -0.233$. Experimentally, the hole-doping is controlled via boron vacancy doping in the substrate\,\cite{ming2023}.

Fig.\,\ref{fig:qpi}\,c shows the QPI data for 10\% hole-doping: the right half of the plot is the original experimental data ($\omega=0$ meV, $T=9$K)\,\cite{ming2023} and the left half contains our theory results (details of theoretical modeling is discussed is in the Supplement\,\cite{supp}). The agreement is excellent, thus we are confident that our non-interacting model in Eq.\,\eqref{ham} correctly describes the metallic regime of hole-doped Sn/Si(111).

{\it Symmetry-broken many-body phases of Sn/Si(111)}.---Electronic correlations are accounted for by adding local and non-local Hubbard interactions, leading to the Hamiltonian
\begin{equation}
    H = H_0 + U \sum_{i} n_{i \uparrow} n_{i \downarrow} + V \sum_{\langle i j \rangle \, \sigma} n_{i \sigma} n_{j \sigma}\ .
\end{equation}
In principle, one might consider longer-ranged repulsions $V_n$, but we assume that the dominant effects are captured by the (bare) onsite $U>0$ and nearest-neighbor $V>0$ Hubbard repulsion. $n^\pd_{i \sigma} = c^\dagger_{i \sigma} c^\pd_{i \sigma}$ is the number operator. The bare interaction parameters can be expressed in terms of the bandwidth of the non-interacting systems, which is $W = 0.5123$eV ($W/t_1= 9.71$).

\begin{figure*}[t]
    \centering
    \includegraphics[width=0.99\textwidth]{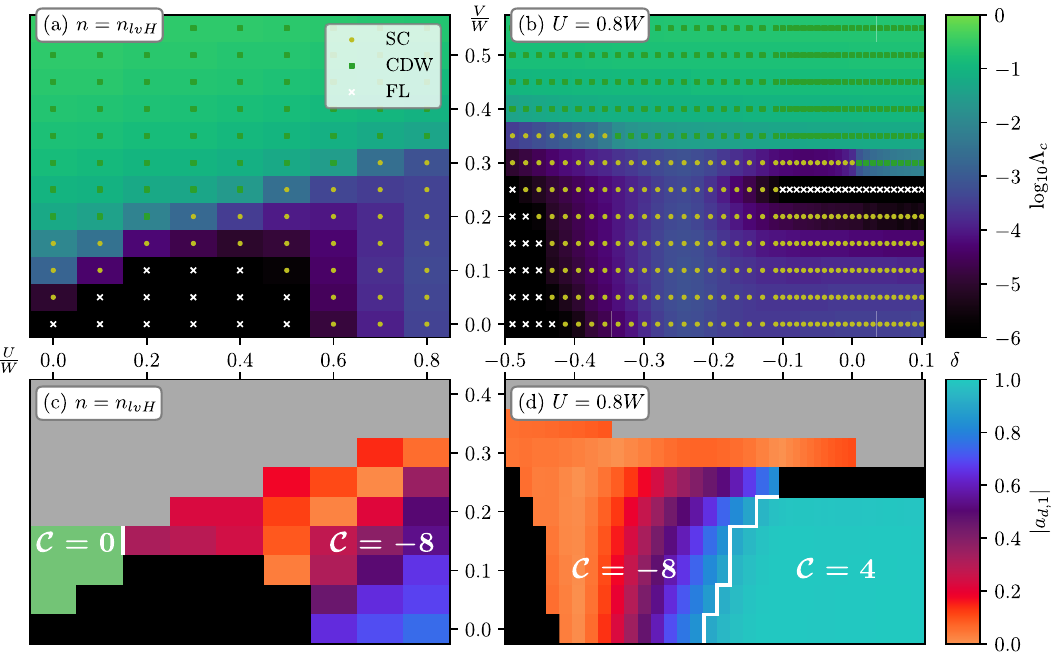}
    \caption{Phase diagrams. (a) $U$-$V$ phase diagram at fixed lower van Hove filling $n_{\text{lvH}} = 0.754 $ ($\delta_{\text {lvH}} = -0.246$). (b) $\delta$-$V$ phase diagram at fixed Hubbard interaction $U = 0.8W$. The color gradient in a and b shows the critical scale $\Lambda_c$ of the divergence in the TUFRG flow. We find three different phases as shown in the legend.
    (c) Combination of the strength of the nearest-neighbor $d$-wave pairing $|a_{d,1}|$ (shown as color gradient) and topological invariant $C$ for the superconducting regime, depending on $U$ and $V$. Green area corresponds to extended $s$-wave superconductivity. Grey (black) area displays charge order (a Fermi liquid). (d) Same as c but depending on $\delta$ and $V$. White line corresponds to a topological phase transition, where the BdG gap closes, from $C=4$ to $C=-8$.
    }
    \label{fig:phase_diags}
\end{figure*}

To compute the low-temperature interacting phases, we perform a perturbative diagrammatic analysis via TUFRG calculations, as implemented in the diverge community code\,\cite{profe2024a, profe2024b}. TUFRG renormalizes the two-particle interaction vertex, interpolating between the bare interactions at high (infinite) temperature and an effective vertex at low temperatures by numerical integration of the flow equation\,\cite{platt2013, metzner2012}. A divergence of the vertex along the path of integration indicates the onset of a symmetry-breaking phase transition, such as superconductivity or charge-order. We then extract the leading instability from the divergent vertex as the leading eigenvalue of the linearized gap equation. In the truncated unity formalism, the instabilities are expressed in terms of the lattice harmonics, providing direct access to the bonding information of the order parameters. For the TUFRG calculations, 61 form factors were included, up to and including a bonding distance of four times nearest neighbors, and the transfer momentum was calculated on a $36^2$ size grid, with a further $51^2$ refinement for loop and loop derivative calculations.

By identifying the many-body instabilities we can construct ground state phase diagrams. Moreover, for superconducting instabilities we can feed the TUFRG order parameter into a Bogoliubov-de Gennes (BdG) mean field theory, giving us access to topological invariants as well as QPI spectra in the superconducting phase.

We show the interacting phase diagrams calculated for fixed doping corresponding to the lower van Hove filling $n_{\text{lvH}} = 0.754$ (Fig.\,\ref{fig:phase_diags}\,a and for fixed $U=0.8 W$ (Fig.\,\ref{fig:phase_diags}\,b.
A white cross indicates that no divergence of the vertex was obtained by the time the scale reached the cutoff of $\Lambda_c = 10^{-5} eV$. This indicates a Fermi liquid phase, which occurs for example when the bare interaction is sufficiently small.

A dominant $V$ drives an instability in the charge channel, with divergences at ordering vectors $\bq = \pm \bK$.  This gives rise to a charge-ordered phase with a $3 \times 3$ reconstructed unit cell (compared to the $\sqrt{3} \times \sqrt{3}$ unit cell of the Sn triangular net, while the Si substrate has $1\times 1$ structure)\,\cite{bunney2026}. This phase has been observed in Pb/Si(111) at half filling, referred to as `1-up 2-down' charge order\,\cite{tresca,adler}. We stress, that for Sn/Si(111) and related compounds this is the only possible $3\times 3$ charge order, unless the aforementioned valley Zeeman term was significantly larger\,\cite{supp}.

Where onsite $U$ dominates, superconductivity emerges. The superconducting critical scale $\Lambda_c$ -- proportional to the critical temperature $T_c$ -- depends on the doping and is largest around the van Hove fillings.
It decreases significantly at larger (more negative) doping and decreases somewhat to the higher, experimentally accessible doping regimes. For dopings around half-filling, the superconducting phase's critical scale decreases with increasing $V$, indicating a competition in the high-temperature Fermi liquid phase between charge order fluctuations and those fluctuations driven by onsite $U$. Our results show the superconducting phase even at half filling; this is based, however, on the assumption that the normal-state is metallic (which is not the case for Sn/Si(111)\,\cite{modesti2007,li2013}--see discussion at end of Letter). The structure of the superconducting order parameter is analyzed further below.

Unlike the superconducting phases, the critical scales calculated across the charge-ordered phase have a weak doping dependence and increase as doping increases. The critical scale also increases with increasing $V$. The structure of the charge instability does not change across either of the two phase diagrams Fig.\,\ref{fig:phase_diags}\,a and b.

Further phase diagrams were calculated: (1) for fixed $U=0.6W$ and (2) for fixed $U=W$, (3) for fixed doping $\delta = -0.2$ (4) for fixed half-filling $\delta = 0$, see Supplement\,\cite{supp}. This additional data demonstrates the stability of the results of Fig.\,\ref{fig:phase_diags} over an extended parameter range.

\begin{figure}[t!]
    \centering
    \includegraphics[width=0.99\columnwidth]{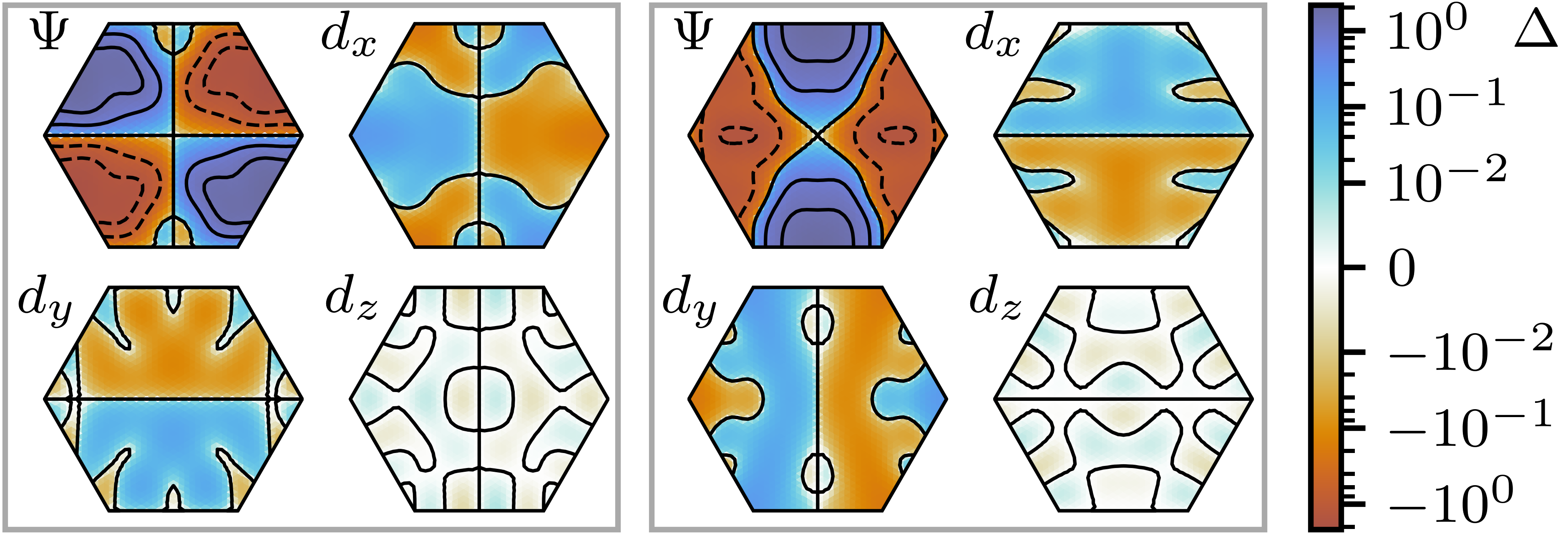}
    \caption{
    Example of the superconducting order parameter in the $C=4$ phase. The two pairings are doubly degenerate. They are decomposed into the spin singlet pairing $\Psi$ and the spin triplet pairings $\boldsymbol{d}$. While the magnitude of $d_x$ and $d_y$ is roughly 10\% of $\Psi$, $d_z$ is smaller by one magnitude. Note the color scale is log normalized to demonstrate this effect.
    Parameters used: $U = 0.8 W$, $V = 0.1W$, $\delta = -0.1$.
        }
    \label{fig:sc_analysis}
\end{figure}

\textit{Superconducting phase analysis}.---
The TUFRG formulation permits the decomposition of the superconducting order parameter, composed of Cooper pairs across lattice bonds, into the different distance pairing shells. For each bonding shell, we then perform a symmetry analysis, where the symmetry group of the Sn adatom layer is $C_{6v}$. (A discussion of the effect of the weak breaking of $C_6$ symmetry due to the substrate is discussed in the Supplement\,\cite{supp}). Significant Rashba spin-orbit coupling couples the point symmetry group (spatial) operations to simultaneous spin-rotations, meaning that the spin structure needs to be taken into consideration when finding symmetry basis functions\,\cite{bunney2024a}. The superconducting instabilities in the phase diagram then form a basis for one of the irreps of $C_{6v}$, which across the calculated phase diagrams is either $E_2$ or $A_1$.

On each pairing shell, there are three possible sets of bonding functions that form an $E_2$ irrep\,\cite{bunney2024a}. Writing the general superconducting order parameter as $ \Delta = ( \Psi + \boldsymbol{d} \cdot \boldsymbol{\sigma} ) \, i \sigma_y \, $\cite{sigrist1991}, the antisymmetric (singlet) spin pairing $\Psi$ has a $d$-wave basis function, and there are two spin symmetric (triplet) pairings $\boldsymbol{d}$ that are $p$- and $f$-wave.
An $E_2$ irrep superconducting order parameter can be a mixture of these three functions spread over a mix of different Cooper pair bonding lengths. However, the $d$-wave pairing dominates the order parameter across the superconducting phase diagrams.
An example set of $E_2$ superconducting order parameters is shown in Fig.\,\ref{fig:sc_analysis}, calculated at 10\% hole doping ($\delta = -0.1$) for bare interaction parameters $U = 0.8 W, V=0.1W$. The doubly degenerate order parameters are compromised primarily of spin-antisymmetric nearest-neighbor $d$-wave pairing. The spin-symmetric $\boldsymbol{d}$ pairings are at most approximately 10\% the amplitude of the spin-singlet (spatial) $d$-wave pairings, roughly matching the ratio of the nearest neighbor Rashba spin-orbit coupling to real valued hopping in the Wannier model.

The primary bonding shell of the $d$-wave pairing shifts across the phase diagram. Fig.\,\ref{fig:phase_diags}\,c and d show the relative weight of the nearest-neighbor $d$-wave pairing, labeled $a_{d,1}$. There is a superconducting phase with nearest-neighbor pairing from van Hove up to half filling, and below van Hove this transitions smoothly to a next-nearest neighbor pairing phase.
This phase transition point is also $V$-dependent, with increased $V$ shifting the transition between the bonding distances to smaller dopings. Increasing $V$ also suppresses the superconducting critical scale $\Lambda_c$ for the nearest-neighbor $d$-wave phase only, which then becomes a small next-nearest neighbor pairing phase around $V/W = 0.3$ before transitioning to the charge ordered phase.
A full bonding shell decomposition, together with details of the explicit calculation of $a_{d,1}$, can be found in the Supplement\,\cite{supp}.

\textit{Topological analysis}.---The complex superposition of the $d$-wave ($E_2$) superconducting pairings, which is energetically most favorable as usual in hexagonal lattice systems\,\cite{honerkamp2003, bunney2024a}, is chiral, \ie it breaks time-reversal symmetry. Thus the  superconducting phases are characterized by the Chern number which can be calculated by forming a mean-field BdG Hamiltonian from the chiral superposition and calculating the Berry curvature numerically over a discretized Brillouin zone\,\cite{fukui2005}. The sum of the Berry curvature of the BdG bands below half-filling is the (band) Chern number\,\cite{bunney2024a}.

The topological superconducting phase diagram is overlaid in Fig.\,\ref{fig:phase_diags}\,c and d. There are two topological phases, a $\mathcal{C}=4$ phase which overlaps with the nearest-neighbor $d$-wave phase, which transitions to a $\mathcal{C} = -8$ phase when the weight of the superconducting order parameter shifts from nearest to next nearest-neighbor $d$-wave pairings.
This change in Chern number can be attributed to the higher harmonics of the next-nearest neighbor pairing, which introduces additional vortices into the superconducting order parameter within the Brillouin zone compared to the nearest neighbor pairing. These additional vortices crossing the Fermi surface forces a gap closing of the BdG spectrum, changing the winding of the overall order parameter along the Fermi surface and therefore the Chern number.
The $s$-wave ($A_1$) superconducting phase does not break time reversal symmetry, and therefore has a topologically trivial superconducting gap ($\mathcal C = 0$).

\begin{figure}
    \centering
    \includegraphics[width=0.7\linewidth]{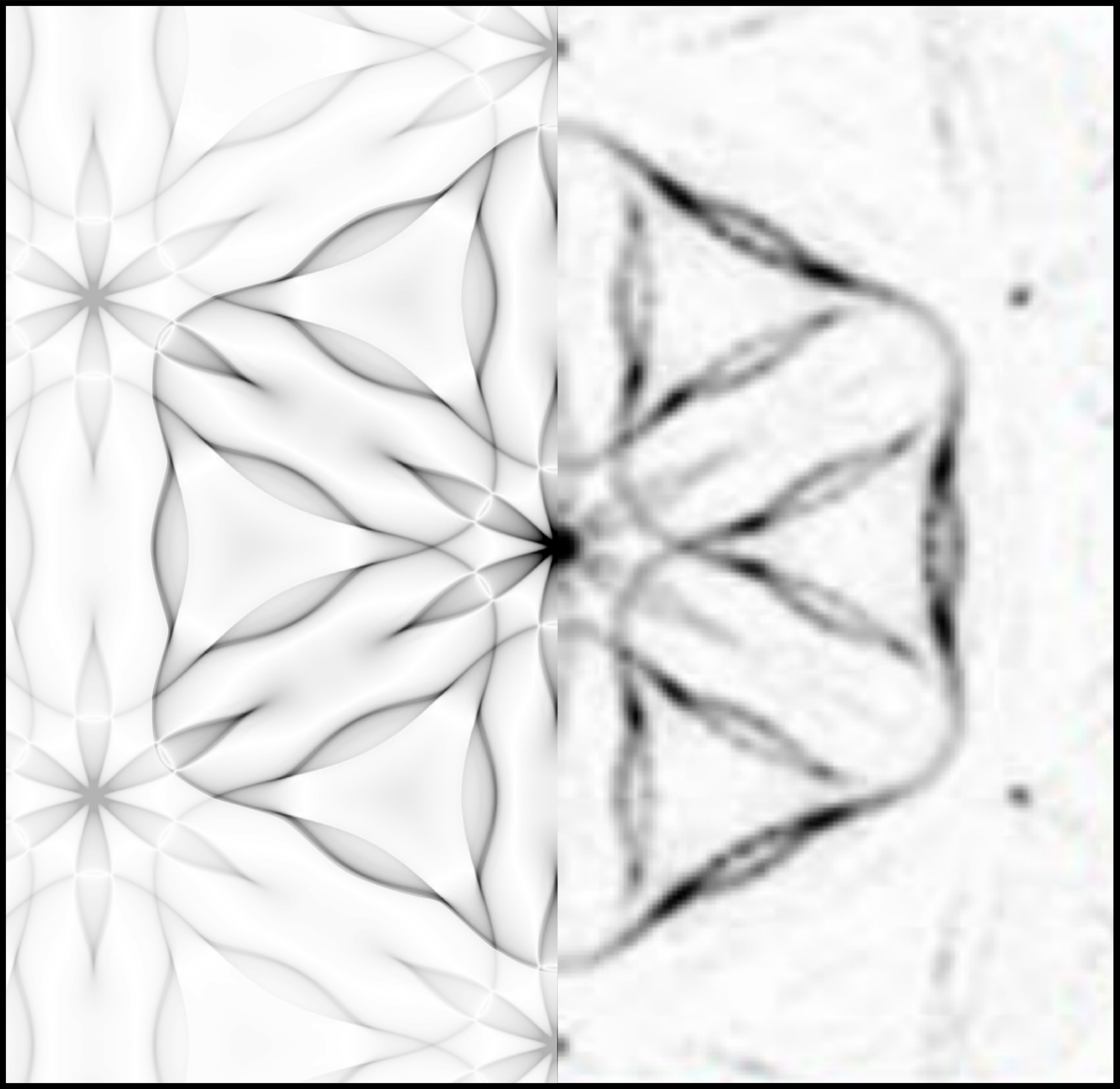}
    \caption{QPI in the superconducting phase: (Right) experimental data\,\cite{ming2023}
    for 10\% hole doping, $\omega=0.8$mV, $T=0.5$K; figure reused with permission of the authors\,\cite{reuse}. (Left) theory for 10\% hole doping ($\delta=0.1$) as described in the main text. The intensity of the theory plot has been reduced outside the main structure to highlight the agreement between experiment and theory.
   }
    \label{fig:qpi_sc}
\end{figure}

\textit{Quasiparticle interference}.---
We now calculate QPI spectra to compare to the experimental QPI scans in the low-temperature superconducting phase.
We introduce a single impurity into the Hamiltonian, and calculate the change in the local density of states compared to the unperturbed system using the Green's function and $T$-matrix formalism.
The QPI spectra are calculated in real space as
\begin{equation} \label{eqn:qpi_paper}
    \delta \rho (\omega, \br) = - \frac{1}{\pi}  \text{Im} [  \text{Tr} [ \hat G_0 (\omega, \br) \, \hat T (\omega) \, \hat G_0 ( \omega, - \br) ] ] \, ,
\end{equation}
with the superconducting Green's function $\hat{G}_0$, $T$-matrix $\hat T ( \omega ) = \hat V[1- \hat G_0 (\omega, \br = 0) \hat V]^{-1}$, and scalar impurity $\hat V = V_0 \hat I$.
The reciprocal space QPI are obtained by Fourier transforming Eq.\,\eqref{eqn:qpi_paper}.
Further details of the QPI calculation are discussed in the Supplement\,\cite{supp}.
We stress that the only free parameters for the QPI calculation at a given bias voltage $\omega$ are the overall superconducting order parameter amplitude $|\Delta|$ and the thermal broadening factor $\Gamma$.  These can be chosen to best fit the calculated QPI spectrum to the experimental data.

Fig.\,\ref{fig:qpi_sc} shows the comparison between our theoretical QPI and the experimental image\,\cite{ming2023}. The theoretical spectrum is calculated for $10 \%$ doping, for $U=0.8W, V=0.1W$, \ie the nearest neighbor $d$-wave phase with a chiral order parameter with $|\Delta| = 2.5$ meV, $\Gamma = 0.01 $ meV, bias voltage $\omega = 0.8$ meV which gives a scalar impurity strength of $V_0 = - 1.24$ eV\,\cite{supp}.
The two spectra share all the common structural features. While some of the features are shared with the metallic QPI shown in Fig.\,\ref{fig:qpi}, the large `flower' or `star'-like structure in the centre of the image only emerges below the superconducting transition temperature, and is near identical between theory and experiment.
In the Supplement\,\cite{supp}, a more detailed QPI analysis is presented, including extended $d$-wave and $s$-wave superconducting order parameters.
Our analysis confirms the earlier finding of Refs.\,\cite{ming2023,wu2025} that the central flower feature appears when time-reversal symmetry is broken, including the topologically trivial $s$-wave pairing in the presence of a magnetic impurity. However, when we calculate (anti-)symmetrized QPI spectra as $\delta \rho^{\pm} (\omega) = ( \delta \rho (\omega) \pm \delta \rho (-\omega) )/2$, we see that the star-like feature remains in both the symmetrized and anti-symmetrized spectra for the topological pairings, but vanishes from the anti-symmetrized spectrum for the trivial $s$-wave pairing, see Suppl.\ Fig.\,S5.

\textit{Discussion and Outlook}. ---
We presented here an in-depth theory analysis of hole-doped Sn/Si(111), substantiating that it is indeed a chiral $d$-wave superconductor.
Experimentally, the half-filled system has been shown to feature a narrow gap from 70K down to 5K\,\cite{modesti2007,li2013}, possibly a Mott gap. We would like to stress that our modeling assumes a metallic normal state, and thus our results at half-filling might not be relevant for Sn/Si(111). Whether this experimental gap is accompanied by charge fluctuations\,\cite{hansmann2016} or magnetic fluctuations\,\cite{profeta2007,li2013} is beyond the scope of this paper. However, it would be very exciting if STM experiments could overcome the difficulty of the resistive substrate and measure $dI/dV$ below 5K. Apart from confirming the (topological) superconducting ground state of hole-doped Sn/Si(111)  by means of other experimental methods in the future, the most exciting avenue is clearly the preparation and measurement of hole-doped samples of Pb/Si(111)\,\cite{tresca,adler}, Sn/SiC(0001)\,\cite{glass-15prl247602} and Pb/SiC(0001)\,\cite{marchetti2025} .

\textit{Acknowledgments}. ---
We acknowledge discussions with R.\ Claessen, R.\ Thomale, L.\ Klebl, E.\ Mascot and H.\ Weitering.
S.R.\ acknowledges support from the Australian Research Council through Grant No.\ DP200101118 and DP240100168.
This research was undertaken using resources from the National Computational Infrastructure (NCI Australia), an NCRIS enabled capability supported by the Australian Government.
This research was supported by The University of Melbourne’s Research Computing Services and the Petascale Campus Initiative.
The authors gratefully acknowledge the scientific support and HPC resources provided by the Erlangen National High Performance Computing Center (NHR@FAU) of the Friedrich-Alexander-Universität Erlangen-N\"urnberg (FAU) under the NHR project ``k102de--FRG". NHR funding is provided by federal and Bavarian state authorities. NHR@FAU hardware is partially funded by the DFG – 440719683.

\bibliography{snsi111_bibfile}

\end{document}


\title{Supplementary Information:\\Chiral topological superconductivity in hole-doped Sn/Si(111)}

\author{Matthew Bunney}
\affiliation{School of Physics, University of Melbourne, Parkville, VIC 3010, Australia}
\affiliation{Institute for Theoretical Solid State Physics, RWTH Aachen University, 52062 Aachen, Germany}
\author{Lucca Marchetti}
\affiliation{School of Physics, University of Melbourne, Parkville, VIC 3010, Australia}
\author{Domenico Di Sante}
\affiliation{Department of Physics and Astronomy, University of Bologna, 40127 Bologna, Italy}
\author{Carsten Honerkamp}
\affiliation{Institute for Theoretical Solid State Physics, RWTH Aachen University, 52062 Aachen, Germany}
\author{Stephan Rachel}
\affiliation{School of Physics, University of Melbourne, Parkville, VIC 3010, Australia}

\date{\today}

\maketitle

\section{Fermi surfaces}

\begin{figure*}[h!]
    \centering
    \includegraphics[width=0.7\linewidth]{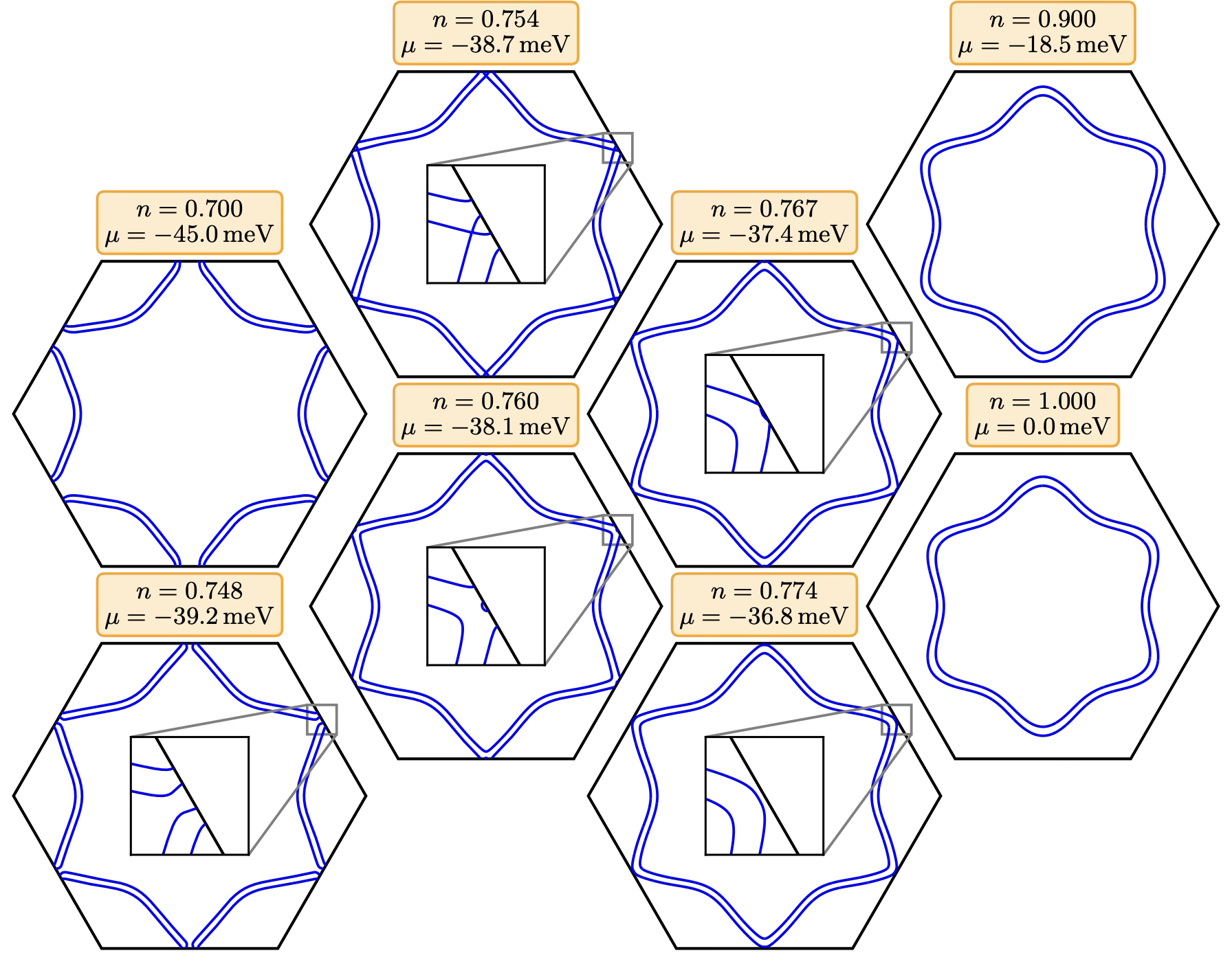}
    \caption{
       Selected Fermi surface plots for Sn/Si(111), based on the Wannier model presented in Ref.\,\cite{marchetti2025}. Increasing in filling from left to right, selected fillings showing the topologically distinct Fermi surfaces, including the two van Hove singularities at $n_{\rm lvH}=0.754$ and $n_{\rm uvH}=0.767$.
    }
    \label{fig:fermi_surfaces}
\end{figure*}

Selected Fermi surfaces for the Wannier model, published in Ref.\,\cite{marchetti2025} and described in Eqn.\,(1), are shown in Fig.\,\ref{fig:fermi_surfaces}.

The Rashba spin-orbit coupling splits the bandstructure, meaning there are two Fermi surfaces and two van Hove singularities. The van Hove singularities split, with a saddle point moving up in energy along the line from $M$ to $K$, and another lowers in energy moving from $M$ to $\Gamma$. The difference in energies is 1.3 meV, which is small compared to the bandwidth $W = 0.5123 eV \, (\Delta \mu_{vH} = 2.54 \times 10^{-3})$.

\newpage

\section{Further phase diagrams}

\begin{figure*}[h!]
    \centering
    \includegraphics[width=0.98\linewidth]{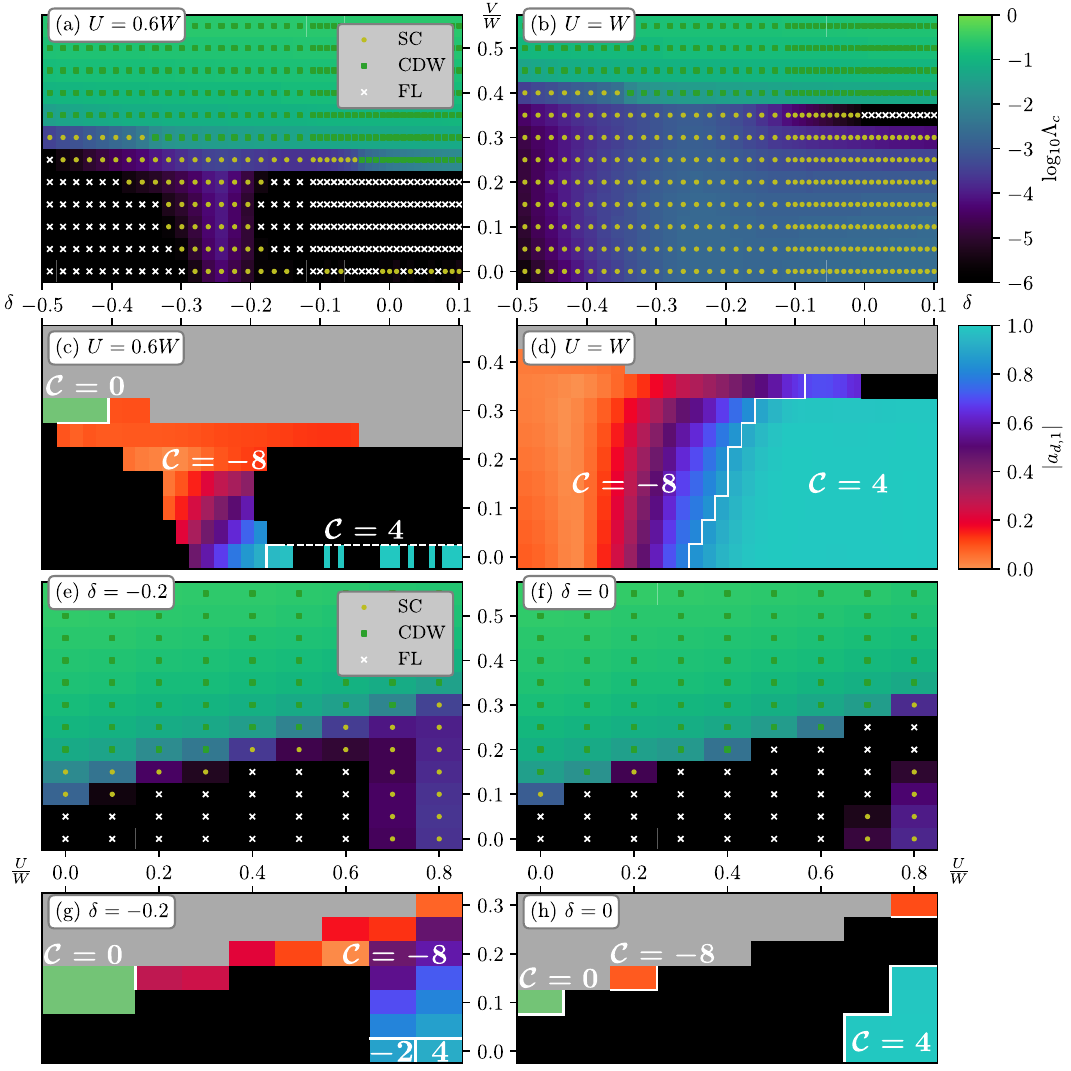}
    \caption{
        Interacting phase diagrams  for fixed (a) $U=0.6W$ and (b) $U=1.0W$ phase diagrams, with corresponding topological superconducting phase diagrams shown in (c) and (d), respectively. Fixed doping phase diagrams are shown for (e) $\delta = -0.2$ and (f) $\delta = 0$, corresponding topological superconducting phase diagrams shown in (g) and (h), respectively.
        The color gradient in the interacting phase diagrams shows the log of the critical scale, with the scattered points indicating the type of FRG divergence/phase.
        In the topological phase diagrams, the color gradient is the nearest-neighbor $d$-wave weight $|a_{d,1}|$ and green areas are (extended) $s$-wave superconducting phases.
    }
    \label{fig:appendix_dopV}
\end{figure*}

As discussed in the main text, further phase diagrams are shown in Fig.\,\ref{fig:appendix_dopV}, calculated for different fixed bare $U = 0.6W$ and $U=W$ as well as fixed dopings $\delta = -0.2$ and $\delta=0$, corresponding to 20\% hole doping and half-filling, respectively.

\begin{figure}[h!]
    \centering
    \includegraphics[width=0.5\linewidth]{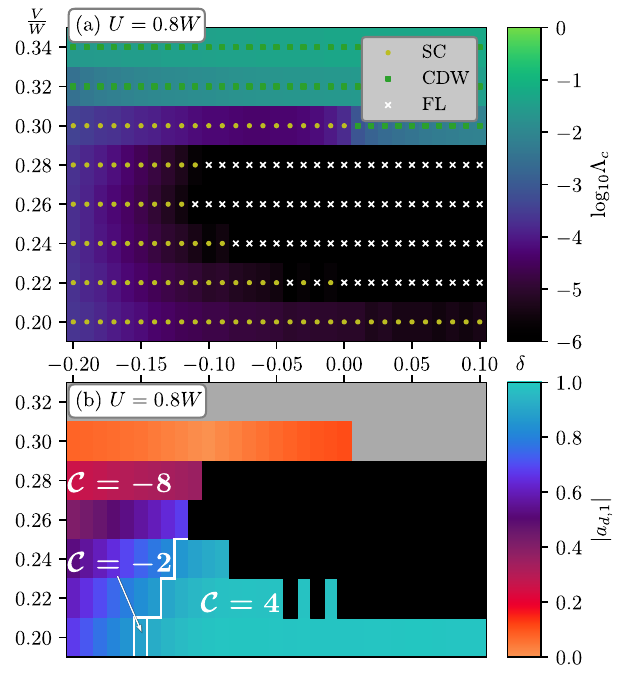}
    \caption{
        (a) Zoomed in interacting phase diagram for fixed $U=0.8W$. The color gradient in the interacting phase diagrams shows the log of the critical scale, with the scattered points indicating the type of FRG divergence/phase. (b) Corresponding topological superconducting phase diagram. The color gradient is the nearest-neighbor $d$-wave weight $|d_{d,1}|$.
    }
    \label{fig:appendix_zoom}
\end{figure}

\newpage
In addition to the phase diagrams above, we show in Fig.\,\ref{fig:appendix_zoom}\,a a zoom-in into the phase diagram Fig.\,2\,b for fixed $U = 0.8W$. It reveals that the Fermi liquid phase separating a superconducting phase a lower and higher $V$ is indeed an extended phase, and not a numerical artifact.
In Fig.\,\ref{fig:appendix_zoom}\,b we show the corresponding nearest-neighbor $d$-wave pairing strength $|a_{d,1}|$ as well as the Chern numbers of the superconducting regime.

\section{Superconducting analysis}

\subsection{\texorpdfstring{$E_2$}{E2} Superconducting Basis Functions}

On each pairing shell $n$, there are three possible sets of bonding functions $\hat \Phi_{i,n}^{E_2}$ that form an $E_2$ irrep\,\cite{bunney2024a}.
\begin{align}
    \hat \Phi_{d,n}^{E_2}:& \; \Psi = \begin{bmatrix}
        \Gamma_{d_{x^2-y^2}} \\ \Gamma_{d_{xy}}
    \end{bmatrix} \, , \label{eqn:d}
    \\
    \hat \Phi_{p,n}^{E_2}:& \; \boldsymbol{d} = \begin{bmatrix}
        ( \Gamma_{p_x}, -\Gamma_{p_y}, 0 ) \\
        ( \Gamma_{p_y}, \Gamma_{p_x}, 0)
    \end{bmatrix} \, ,
    \\
    \hat \Phi_{f,n}^{E_2}:& \; \boldsymbol{d} = \begin{bmatrix}
        ( \Gamma_{f}, 0, 0 ) \\
        ( 0, \Gamma_{f}, 0)
    \end{bmatrix} \, , \label{eqn:f}
\end{align}
where
\begin{equation}
    \hat \Delta = ( \Psi + \boldsymbol{d} \cdot \boldsymbol{\hat \sigma} ) \, i \hat \sigma_y
\end{equation}
is the conventional breakdown of pairings in spin antisymmetric and symmetric components (under exchange of spin indices of the Cooper pairs), with the Pauli basis $\hat \sigma$ acting on the spin index. Note that the hat indicates that it is a matrix in spin space. The three possible basis functions shown in Eqns.\,\eqref{eqn:d}-\eqref{eqn:f} are expressed in terms of $d$, $p$ and $f$-wave pairings, respectively, which indicates their angular momentum winding around the $\Gamma$ point (in reciprocal space).
An $E_2$ irrep superconducting order parameter can therefore be a mix of these three functions spread over a mix of different Cooper pair bonding lengths, which can be expressed as
\begin{equation}
    \hat \Delta = \sum_{i,n} a_{i,n} \hat \Phi_{i,n}^{E_2} \, .
\end{equation}
The $E_2$ superconducting order parameter can be fully determined by the coefficients $a_{i,n}$.

\begin{figure*}[h!]
    \centering
    \includegraphics[width=0.95\linewidth]{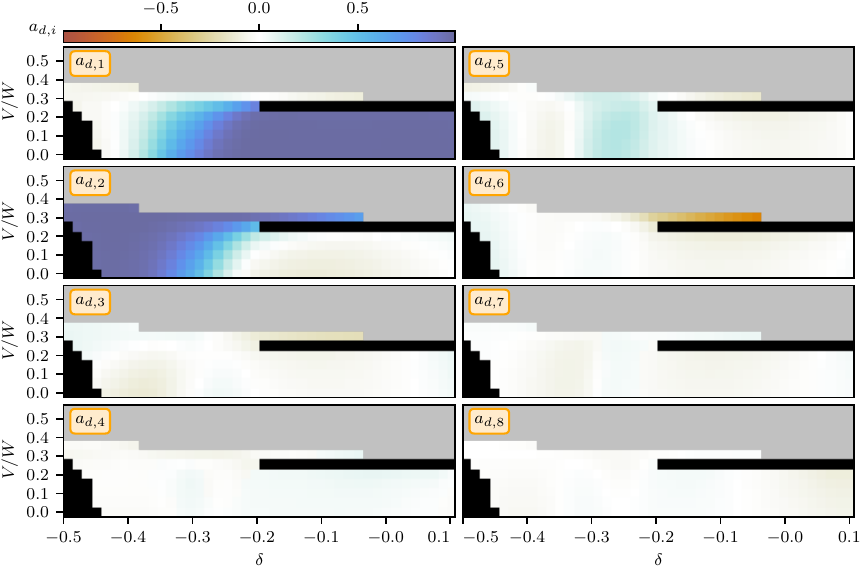}
    \caption{
    Basis function breakdown for the interacting phase diagram at fixed $U=08.W$, showing the $E_2$ irrep $d$-wave basis functions across the superconducting phase.}
    \label{fig:irreps_U08}
\end{figure*}

Fig.\,\ref{fig:irreps_U08} shows the breakdown of the $d$-wave order parameters over the different pairing lengths included in the TUFRG calculation, for the interacting phase diagram shown in Fig.\,2 b and d. The coefficients $a_{d,n}$ on the different pairing shells $(n)$ are calculated as:
\begin{equation}
    a_{d,n} = \sum_{\boldsymbol{k}} \text{Tr}  \left[ (\hat{\Phi}_{d,n}^{E_2})^{\dagger} \hat{\Delta} \, \right] \, .
\end{equation}
As mentioned in the main text, the spin-triplet superconducting weights are at most approximately 10\% of the $d$-wave weights on the same pairing shell. We can see the main physics is captured on nearest and next-nearest neighbors, with the exception of a considerable sixth nearest neighbor pairing in the superconducting phase for $V=0.3W$, near the charge ordered phase. As demonstrated in Fig.\,\ref{fig:appendix_zoom}, it is not clear how extensive across parameter space this superconducting phase is.

\subsection{\texorpdfstring{$C_6$}{C6} symmetry breaking}
While the symmetry group of the Sn layer is $C_{6v}$, there is a small $C_6$ rotation symmetry breaking effect of the substrate in the Wannier model, of the size $ \sim 0.01 t_1$\,\cite{marchetti2025}. The effect of this on the superconducting order parameter is to allow the mixing of $E_1$ and $E_2$ order parameters, since they are both basis functions of the $E$ irrep of $C_{3v}$. The only possible $E_1$ order parameter is
\begin{equation}
    \Gamma_{n}^{E_1}: \; \boldsymbol{d} = \begin{bmatrix}
    ( 0, 0, \Gamma_{p_x} ) \\
    ( 0, 0, \Gamma_{p_y})
    \end{bmatrix}\ .
\end{equation}
Since this is the only $E$ irrep superconducting order parameter with a non-zero $d_z$ component, this symmetry breaking can be seen directly in the order parameter in Fig.\,3 with the distinct (extended) $p$-wave pattern in the $d_z$ regime. These $E_1$ irrep SC components are small, on the same order as the symmetry breaking in the normal Hamiltonian, so therefore would not have a large or measurable effect on the physics of the superconducting phases.

\section{Quasi-Particle Interference}
We consider a local impurity in the presence of the superconducting phase defined by the Hamiltonian
\begin{equation}
    H = H_0 + V \, ,
\end{equation}
where $H_0$ is the unperturbed (impurity-free) superconducting Hamiltonian. $V$ is the local impurity term, given as
\begin{equation} \label{eqn:impurity_ham}
    V = \sum_{i \, \sigma \sigma'} [ \hat V ]_{\sigma \sigma'}^{\phantom{\dagger}} \delta^{\phantom{\dagger}}_{i, \, \boldsymbol{R} = 0} c^\dagger_{i \sigma} c^{\phantom{\dagger}}_{i \sigma'} \, ,
\end{equation}
where $\hat V = V_0 \, \hat I$ for a scalar impurity and $\hat V = V_0 \, \hat \sigma_z$ for a magnetic impurity\,\cite{balatsky2006}.
The QPI spectra are calculated using the $T$-matrix formalism\,\cite{hirschfeld2015, bunney2026},
\begin{equation} \label{eqn:qpi_rs}
    \delta \rho (\omega, \br) = - \frac{1}{\pi}  \text{Im} [  \text{Tr} [ \hat G_0 (\omega, \br) \, \hat T (\omega) \, \hat G_0 ( \omega, - \br) ] ] \, ,
\end{equation}
where the $T$-matrix can be shown to take the form:
\begin{equation}
    \hat T ( \omega ) = \frac{\hat V}{1- \hat G_0 (\omega, \br = 0) \hat V} \; .
\end{equation}
While this definition of the $T$-matrix is a formal definition, in practice it can be calculated as the non-trivial solution to the matrix equation
\begin{equation}
    (1- \hat G_0 (\omega, \br = 0) \hat V) \, \hat T ( \omega ) = \hat V\ .
\end{equation}
Note that the Born approximation $\hat T = \hat V$ corresponds to a first-order approximation of the binomial expansion of the $T$-matrix.
Subgap states (that exist in the superconducting gap) are associated with the poles of the $T$-matrix (at zero temperature)\,\cite{callaway2013}. The condition for the existence of these bound states is
\begin{equation} \label{eqn:bound_evals}
    \text{det} [ 1 - \hat G_0 (\omega, \br = 0) \, \hat V ] = 0 \, .
\end{equation}
Assuming a given unperturbed Green's function (Hamiltonian) and $\omega$, this fixes the value of $V_0$ in Eqn.\,\eqref{eqn:impurity_ham}. Note that in principle this could introduce large $V_0$ values, so accuracy could potentially be improved by including bands higher in energy now accessible to the impurity states.

The QPI modulation shown in Fig.\,1\,c and Fig.\,4 is then the Fourier transform of Eqn.\,\eqref{eqn:qpi_rs},
\begin{equation}
    \delta \rho (\bq) = \int \frac{d \br}{(2\pi)^2} e^{i \br \cdot \bq} \, \delta \rho (\br)\ .
\end{equation}
A flat cutoff is introduced to remove the sharp peak at $\bq=0$, similar to how the experimental data is processed.

\begin{figure*}[p]
    \centering
    \includegraphics[width=0.95\linewidth]{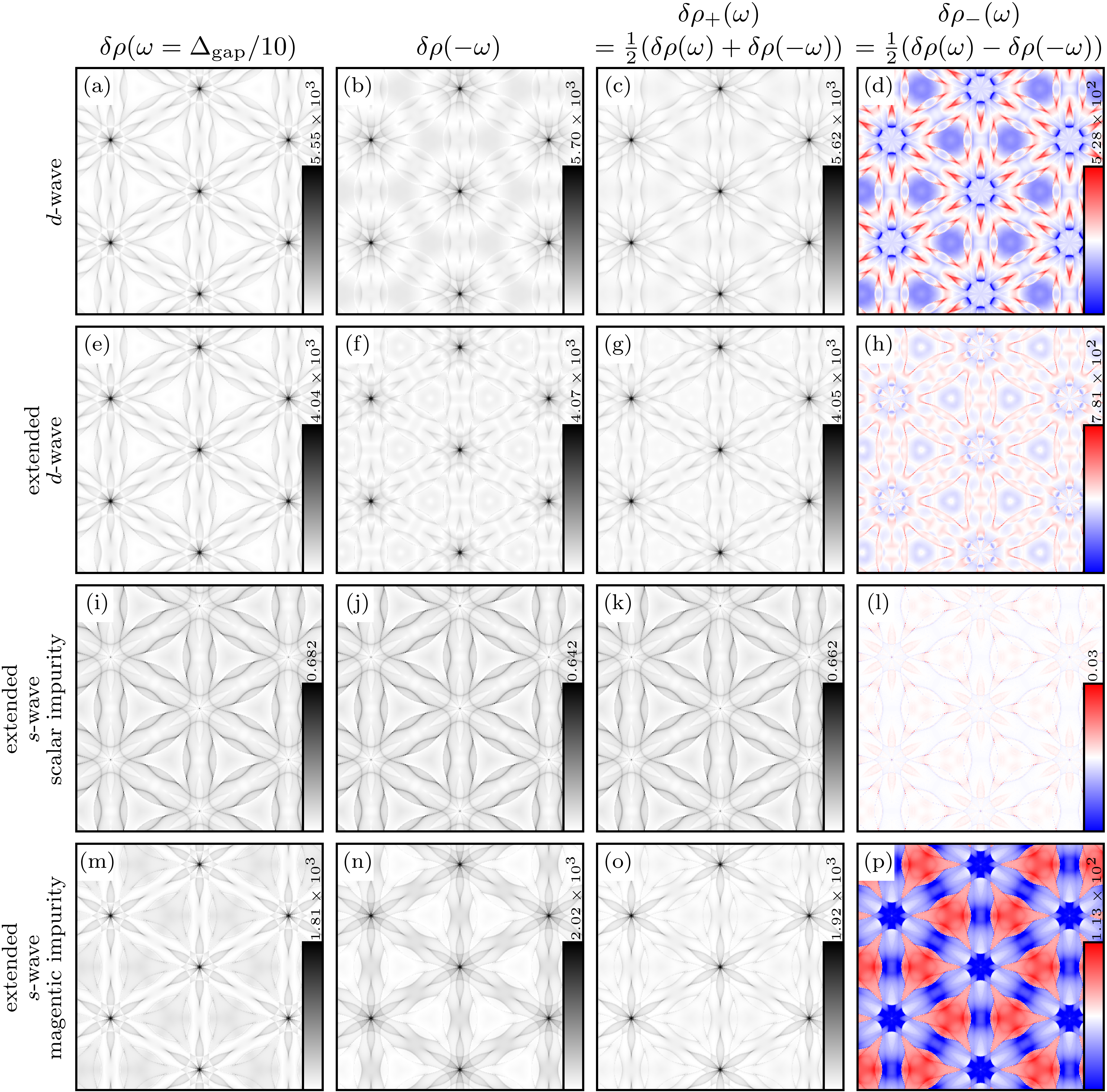}
    \caption{
        Simulated QPI spectra for the different pairings found in the interacting phase diagram; (a-d) shows QPI for a $d$-wave pairing, (e-h) extended (next-nearest neighbor) $d$-wave pairing, (i-l) extended $s$-wave pairing with a scalar impurity and (m-p) extended $s$-wave pairing with a magnetic impurity. Note that the scalar and magnetic cases for both $d$-wave pairings look nearly identical, so only the scalar impurity case is shown. All figures were calculated with the same order parameter amplitude $|\Delta| = 10^{-3}$ meV and with thermal broadening factor $\Gamma = 10^{-5}$ meV. For each the pairings, BdG gap $\Delta_{\text{gap}}$ is calculated to be $ 1.09 \times 10^3$ meV, $3 \times 10^{-4}$ meV and $4.34 \times 10^{-4}$ meV for $d$-wave, extended (next-nearest neighbor) $d$-wave and extended-$s$ wave (nearest neighbor) pairings respectively. The strength of the impurities $V_0 = 0.950$ meV for the $d$-wave pairing, $0.752$ meV for the extended $d$-wave pairing, and $-0.135$ meV for a scalar impurity with the extended $s$-wave and $0.112$ meV for the magnetic impurity with extended $s$-wave.}
    \label{fig:qpi_grid}
\end{figure*}

In Fig.\,\ref{fig:qpi_grid} we show additional calculated QPI plots for both scalar and magnetic impurities as well as for extended $d$-wave and extended $s$-wave superconducting order parameters, in addition to the nearest-neighbor $d$-wave order parameter shown in the main text. These superconducting instabilities appear in the phase diagrams all at experimentally attainable dopings, \textit{e.g.}, see Fig.\,\ref{fig:appendix_dopV} g and h. The nearest neighbor $d$-wave pairing order parameter was the resulting FRG instability with parameter values $U = 0.8W, V = 0.1W, \delta = -0.1$, the extended $d$-wave pairing was calculated at $U = 0.8W, V = 0.35W, \delta = -0.1$ and the extended $s$-wave pairing was calculated at $U=0, V=0.1W,\delta=-0.2$. Note that in the QPI spectra, the $d$-wave pairings are the chiral superposition of the degenerate $E_2$ instabilities, which is what minimizes the free energy\,\cite{bunney2026}.

QPI spectra were calculated for both scalar and magnetic impurities. For the (extended) $d$-wave pairings, there was no difference in the magnetic and scalar impurity cases, hence the magnetic impurities have been omitted and only the scalar impurities are shown. We notice that while the agreement for the $d$-wave case is excellent, similar QPI spectra can be simulated using the extended $d$-wave pairing, and extended $s$-wave pairing with a magnetic impurity. In order to distinguish between the pairings, we turn the the (anti-)symmetrized QPI spectra, given by the (anti-)symmetrized modulation of the local density of state
$\delta \rho^{+}$ ($\delta \rho^{-}$):
\begin{equation}
    \delta \rho^{\pm} (\omega) = \frac{1}{2} ( \delta \rho (\omega) \pm \delta \rho (-\omega)  )\ ,
\end{equation}
which exploits particle-hole symmetry of the BdG eigenstates to isolate the parts of the spectra that depend on the phase dependence of the superconducting pairing\,\cite{hirschfeld2015, altenfeld2018, boker2020}. In particular, this method has been applied to superconducting systems such as Sr$_2$RuO$_4$\,\cite{du2018, profe2024c} and has been suggested as a method of distinguishing the spin triplet order parameter of the high-field superconductor UTe$_2$\,\cite{christiansen2025}.

The first and second columns of Fig.\,\ref{fig:qpi_grid} show pairs of QPI spectra at bias voltages $\omega$ and $-\omega$, respectively. The value of $\omega = \Delta_{\text{gap}}/10$ was chosen to match the experimental data (presented on the right of Fig.\,4), where the central dark star feature was most prominent within the superconducting gap. In our model, $\Delta_{\text{gap}}$ was calculated as minimum energy of the top BdG bands, where the superconducting order parameter amplitude was set to $\Delta = 10^{-3}$ meV. This matches the experimental gap observed in differential conductance maps\,\cite{ming2023}.
The third and fourth columns of Fig.\,\ref{fig:qpi_grid} show the symmetrized and anti-symmetrized QPI, respectively. While the central dark star remains a feature of the symmetrized QPI for the $d$-wave pairings, and the extended $s$-wave with the magnetic impurity, a discernible star-like feature only appears in the anti-symmetrized pairings for the $d$-wave pairings, and vanishes for the magnetic impurity extended $s$-wave QPI.
The extended $s$-wave scalar impurity case is the opposite to the magnetic one: while the symmetrized QPI has no star feature -- like the calculated spectra at bias voltage $\omega$ -- the anti-symmetrized spectrum has a faint star feature.
This analysis leads us to contend that the signature of the topological chiral $d$-wave pairing is not necessarily just the presence of the dark central star-like feature, but rather its persistence through to both the symmetrized and anti-symmetrized QPI spectra.

\newpage
\afterpage{\clearpage}

\newpage

\bibliography{snsi111_bibfile}